 \documentclass[twocolumn]{aastex6}

\usepackage[latin1]{inputenc}

\def\kms    {\ifmmode{{\rm \ts km\ts s}^{-1}}\else{\ts km\ts s$^{-1}$}\fi}
\def\msol   {\ifmmode{{\rm M}_{\odot} }\else{M$_{\odot}$}\fi}
\def\lsol   {\ifmmode{L_{\odot}}\else{$L_{\odot}$}\fi}
\def\lfir   {\ifmmode{L_{\rm FIR}}\else{$L_{\rm FIR}$}\fi}
\def\lir   {\ifmmode{L_{\rm IR}}\else{$L_{\rm IR}$}\fi}
\def\zsol   {\ifmmode{{\rm Z}_{\odot}}\else{Z$_{\odot}$}\fi}
\def\etal   {{\rm et\ts al.}}
\def\tkin   {\ifmmode{T_{\rm kin}}\else{$T_{\rm kin}$}\fi}
\def\hh   {\ifmmode{{\rm H}_{2}}\else{H$_{2}$}\fi}

\def\ts     {\thinspace}
\def\ci     {\ifmmode{[{\rm C}{\rm \small I}]}\else{[C\ts {\scriptsize I}]}\fi}
\def\cii    {\ifmmode{[{\rm C}{\rm \small II}]}\else{[C\ts {\scriptsize II}]}\fi}
\def\oiii    {\ifmmode{[{\rm O}{\rm \small III}]}\else{[O\ts {\scriptsize III}]}\fi}
\def\nii    {\ifmmode{[{\rm N}{\rm \small II}]}\else{[N\ts {\scriptsize II}]}\fi}
\def\mue {\ifmmode{\mu{\rm m}}\else{$\mu$m}\fi}

\usepackage{color}

\definecolor{purple}{rgb}{0.6, 0.4, 0.8}

\begin{document}


\title{ISM properties of a Massive Dusty Star-Forming Galaxy discovered at $z \sim 7$}


\def\MPIfR{1}
\def\IMPRS{2}
\def\ESOGarching{3}
\def\Arizona{4}
\def\IllinoisAstr{5}
\def\IllinoisPhys{6}
\def\Diego{7}
\def\CfA{8}
\def\Marseille{9}
\def\Cambridge{10}
\def\Kavli{11}
\def\JPL{12}
\def\KICPChicago{13}
\def\PhysicsUChicago{14}
\def\EFIChicago{15}
\def\AAUChicago{16}
\def\Dal{17}
\def\Mary{18}
\def\Davis{19}
\def\UFlorida{20}
\def\UCL{21}
\def\Durham{22}
\def\CompAstro{23}
\def\Stanford{24}
\def\Hubble{25}
\def\UCLA{26}
\def\NRAO{27}

\author{
M.~L.~Strandet\altaffilmark{\MPIfR,\IMPRS},
%
%
A.~Wei\ss\altaffilmark{\MPIfR},
C.~De~Breuck\altaffilmark{\ESOGarching},
D.~P.~Marrone\altaffilmark{\Arizona},  
J.~D.~Vieira\altaffilmark{\IllinoisAstr,\IllinoisPhys},
%
M.~Aravena\altaffilmark{\Diego},
M.~L.~N.~Ashby\altaffilmark{\CfA},
M.~B\'ethermin\altaffilmark{\Marseille},
M.~S.~Bothwell\altaffilmark{\Cambridge,\Kavli},
C.~M.~Bradford\altaffilmark{\JPL},
J.~E.~Carlstrom\altaffilmark{\KICPChicago,\PhysicsUChicago,\EFIChicago,\AAUChicago}, 
S.~C.~Chapman\altaffilmark{\Dal},
D.~J.~M.~Cunningham\altaffilmark{\Dal, \Mary},
Chian-Chou~Chen\altaffilmark{\ESOGarching},
C.~D.~Fassnacht\altaffilmark{\Davis},
A.~H.~Gonzalez\altaffilmark{\UFlorida}, 
T.~R.~Greve\altaffilmark{\UCL},	
B.~Gullberg\altaffilmark{\Durham}, 
C.~C.~Hayward\altaffilmark{\CfA,\CompAstro}, 
Y.~Hezaveh\altaffilmark{\Stanford,\Hubble},
K.~Litke\altaffilmark{\Arizona},
J.~Ma\altaffilmark{\UFlorida}, 
M.~Malkan\altaffilmark{\UCLA},
K.~M.~Menten\altaffilmark{\MPIfR},
T.~Miller\altaffilmark{\Dal},
E.~J.~Murphy\altaffilmark{\NRAO},
D.~Narayanan\altaffilmark{\UFlorida}, 
K.~A.~Phadke\altaffilmark{\IllinoisAstr},
K.~M.~Rotermund\altaffilmark{\Dal},
J.~S.~Spilker\altaffilmark{\Arizona},
J.~Sreevani\altaffilmark{\IllinoisAstr}
}

\altaffiltext{\MPIfR}{Max-Planck-Institut f\"{u}r Radioastronomie, Auf dem H\"{u}gel 69 D-53121 Bonn, Germany}
\altaffiltext{\IMPRS}{Member of the International Max Planck Research School (IMPRS) for Astronomy and Astrophysics at the Universities of Bonn and Cologne}
\altaffiltext{\ESOGarching}{European Southern Observatory, Karl Schwarzschild Stra\ss e 2, 85748 Garching, Germany}
\altaffiltext{\Arizona}{Steward Observatory, University of Arizona, 933 North Cherry Avenue, Tucson, AZ 85721, USA}
\altaffiltext{\IllinoisAstr}{Department of Astronomy, University of Illinois, 1002 West Green St., Urbana, IL 61801}
\altaffiltext{\IllinoisPhys}{Department of Physics, University of Illinois, 1002 West Green St., Urbana, IL 61801}
\altaffiltext{\Diego}{N\'ucleo de Astronom\'{\i}a, Facultad de Ingenier\'{\i}a y Ciencias, Universidad Diego Portales, Av. Ej\'ercito 441, Santiago, Chile}
\altaffiltext{\CfA}{Harvard-Smithsonian Center for Astrophysics, 60 Garden Street, Cambridge, MA 02138, USA}
\altaffiltext{\Marseille}{Aix Marseille Univ, CNRS, LAM, Laboratoire d'Astrophysique de Marseille, Marseille, France}
\altaffiltext{\Cambridge}{Cavendish Laboratory, University of Cambridge, 19 J.J. Thomson Avenue, Cambridge, CB3 0HE, UK}
\altaffiltext{\Kavli}{Kavli Institute for Cosmology, University of Cambridge, Madingley Road, Cambridge CB3 0HA, UK}
\altaffiltext{\JPL}{Jet Propulsion Laboratory, 4800 Oak Grove Drive, Pasadena, CA 91109, USA}
\altaffiltext{\KICPChicago}{Kavli Institute for Cosmological Physics, University of Chicago, 5640 South Ellis Avenue, Chicago, IL 60637, USA}
\altaffiltext{\PhysicsUChicago}{Department of Physics, University of Chicago, 5640 South Ellis Avenue, Chicago, IL 60637, USA}
\altaffiltext{\EFIChicago}{Enrico Fermi Institute, University of Chicago, 5640 South Ellis Avenue, Chicago, IL 60637, USA}
\altaffiltext{\AAUChicago}{Department of Astronomy and Astrophysics, University of Chicago, 5640 South Ellis Avenue, Chicago, IL 60637, USA}
\altaffiltext{\Dal}{Dalhousie University, Halifax, Nova Scotia, Canada}
\altaffiltext{\Mary}{Department of Astronomy and Physics, Saint Mary's University, Halifax, Nova Scotia, Canada}
\altaffiltext{\Davis}{Department of Physics,  University of California, One Shields Avenue, Davis, CA 95616, USA}
\altaffiltext{\UFlorida}{Department of Astronomy, University of Florida, Bryant Space Sciences Center, Gainesville, FL 32611 USA}
\altaffiltext{\UCL}{Department of Physics and Astronomy, University College London, Gower Street, London WC1E 6BT, UK}
\altaffiltext{\Durham}{Centre for Extragalactic Astronomy, Department of Physics, Durham University,  South Road,  Durham DH1 3LE,  UK}
\altaffiltext{\CompAstro}{Center for Computational Astrophysics, Flatiron Institute, 162 Fifth Avenue, New York, NY 10010, USA}
\altaffiltext{\Stanford}{Kavli Institute for Particle Astrophysics and Cosmology, Stanford University, Stanford, CA 94305, USA}
\altaffiltext{\Hubble}{Hubble Fellow}
\altaffiltext{\UCLA}{Department of Physics and Astronomy, University of California, Los Angeles, CA 90095-1547, USA}
\altaffiltext{\NRAO}{National Radio Astronomy Observatory, 520 Edgemont Road, Charlottesville, VA 22903, USA}

\begin{abstract}
We report the discovery and constrain the physical conditions of the interstellar medium of the highest-redshift millimeter-selected dusty star-forming 
galaxy (DSFG) to date, SPT-S J031132$-$5823.4 (hereafter SPT0311$-$58), at $z=6.900 \pm 0.002$.  SPT0311$-$58 was discovered via its 1.4\,mm thermal dust continuum emission in the South Pole Telescope (SPT)-SZ survey.  The spectroscopic redshift was determined through an ALMA 3\,mm frequency scan that detected CO(6--5), CO(7--6) and \ci(2--1), and subsequently confirmed by detections of CO(3--2) with ATCA and \cii\ with APEX.  
We constrain the properties of the ISM in SPT0311$-$58 with a radiative transfer analysis of the 
dust continuum photometry and the CO and \ci\ line emission.  This allows us to determine the gas content without \emph{ad hoc} assumptions about gas mass scaling factors.  
SPT0311$-$58 is extremely massive, with an intrinsic gas mass of $M_{\rm gas} = 3.3 \pm 1.9 \times10^{11}\,\msol$. 
Its large mass and intense star formation is very rare for a source well into the Epoch of Reionization.
\end{abstract}


\section{Introduction}
Searches for the most distant galaxies have now reached as far back as the first billion years in the history of the Universe, and are peeking into the epoch of reionization (EoR) at $6<z<11$ \citep{planck16}.  Some of the most important questions in observational cosmology concern the time scale over which the reionization of the Universe took place, the identification of the objects providing the ionizing photons and the enrichment of galaxies with metals.
It is expected that star-forming galaxies play a major role in the reionization, so to understand the evolution of the Universe from its neutral beginning to its present ionized state we must study the galaxies in the EoR \citep[see reviews by][]{stark16,bouwens16}. How galaxies formed and evolved in the EoR is unknown. Galaxies in this era are currently being found from rest-frame ultraviolet (UV) surveys \citep[e.g.][]{ouchi10}.
Most of these systems, however, are low-mass star-forming galaxies for which the enrichment of the cold ISM is difficult to study even in long integrations with the Atacama Large Millimeter/submillimeter Array (ALMA) \citep[][b]{bouwens16b}.

Massive dusty star-forming galaxies (DSFGs) \citep{casey14} are not expected to be found into the EoR 
because it is difficult to produce their large dust masses within a few hundred Myr of the Big Bang \citep[][]{ferrara10,mattsson15}. 
Recent wide-area {\it Herschel} and optical QSO surveys, however, have revealed dusty galaxies out to $z\sim6-7$ \citep[e.g.,][]{venemans12,riechers13}.
These systems offer the unique opportunity to study extreme cases of metal/dust enrichment of the ISM within the EoR in the most massive over-densities at these redshifts.

Here we present the DSFG SPT-S J031132$-$5823.4 (hereafter SPT0311$-$58) discovered in the South Pole Telescope (SPT)-SZ survey \citep{carlstrom11, vieira13}. 
SPT0311$-$58 is the highest redshift millimeter-selected DSFG known to-date, located well into the EoR at a redshift of $z=6.900 \pm 0.002$.  With this source, we take a step of almost 100\,Myr closer to the Big Bang than the previously most distant DSFG at $z=6.34$ found by \citet{riechers13}, bringing us $\sim$760\,Myr away from Big Bang.
Throughout the paper, we assume a $\Lambda$CDM cosmology with H$_{0}$=70\,\kms\ Mpc$^{-1}$, $\Omega_\Lambda$\,=\,0.7 and $\Omega_{\rm M}$\,=\,0.3.

\section{Observational Results} 
\label{sec:observations}
\subsection{Determining the redshift}
\label{alma3mm}

\begin{figure*}
\includegraphics[viewport= 100 498 510 730, clip=true,width= \textwidth,angle=0]{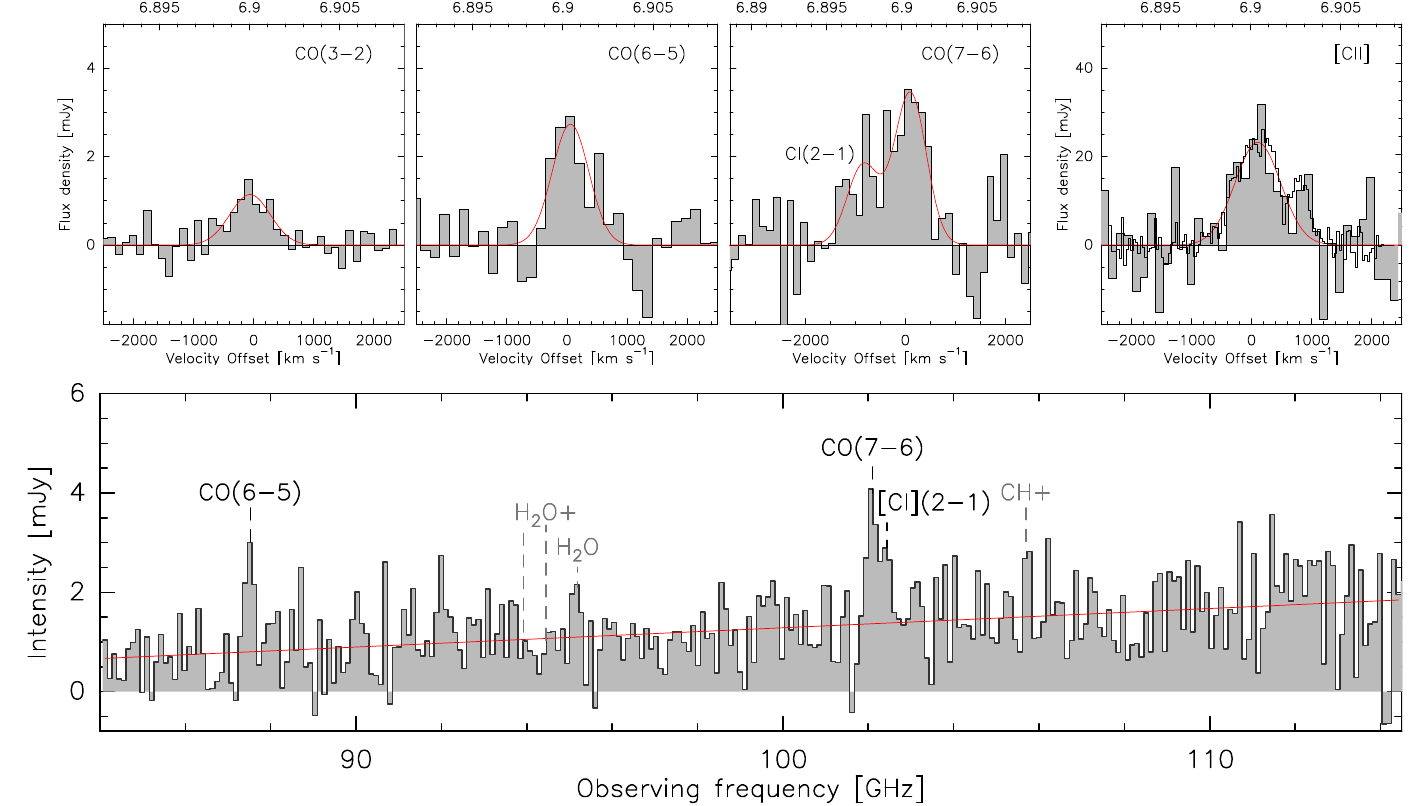}
\caption{\small The lower part of the figure shows the ALMA 3\,mm spectrum of SPT0311$-$58 spanning 84.2 -- 114.9\,GHz. 
The spectrum has been binned to best show the lines. Transitions labeled in black are detected and grey labels indicate where other transitions should be. The red line indicate the zeroth order baseline.
The sub-panels above the spectrum show, from left to right, the continuum-subtracted spectra of:  ATCA CO(3--2), ALMA CO(6--5), ALMA CO(7--6) and CI(2--1), and APEX \cii\ with ALMA \cii\ overlaid as a solid black histogram. Gaussian fits to the spectra are shown in red.
}
\label{Fig:spectra}
\end{figure*}

The redshift search for SPT0311-58 was performed in ALMA band 3 by combining five tunings covering 84.2 -- 114.9\,GHz \citep[project ID:
2015.1.00504.S; see][for further details on the observing setup]{weiss13,strandet16}.  The observations were carried out on 2015 December 28 and 2016 January 2 in the Cycle 3 compact array configuration.  The number of antennas varied from 34 to 41, with baselines up to 300\,m yielding a synthesized beam size of $2\farcs 2 - 3\farcs 0$.  Typical system temperatures for the observations were $T_{\rm sys}$ = 50 -- 80\,K (SSB).  Flux calibration was done with Uranus, bandpass calibration with J0334$-$4008, and phase calibration with J0303$-$6211 and J0309$-$6058.  The on-source time varied between 60 seconds and 91 seconds per tuning, accounting for a total of 6 minutes and 10 seconds.  The data were processed using the Common Astronomy Software Application package \citep[CASA,][]{mcmullin07}.

We created a cleaned 3\,mm continuum image combining all 5 tunings.  This yields a high signal to noise ratio (SNR) detection of $\sim$35. 
We also generated a spectral cube using natural weighting with a channel width of 19.5\,MHz (50 -- 65\,kms$^{-1}$ for the highest and lowest observing frequency, respectively), which gives a typical noise per channel of 0.9 --1.7\,mJy beam$^{-1}$.

The ALMA 3\,mm spectrum of SPT0311$-$58 was extracted at the centroid of the 3\,mm continuum emission ($\alpha$: 03$^{\rm h}$11$^{\rm m}$33$\fs$142 $\delta$: $-58^{\circ}$23$'$33$\farcs$37 (J2000)) and is shown in Figure \ref{Fig:spectra}. We detect emission in the CO $J$ = 6 -- 5 and 7 -- 6 lines and the \ci\ $^3P_2-^3P_1$ line (in the following $2-1$) 
and their noise-weighted line frequencies yield a redshift of $z=6.900 \pm 0.002$. We also see hints of H$_2$O($2_{11}$ -- $2_{02}$) and CH$^+$(1--0), but these are not formally detected in this short integration. 

The line and continuum properties are given in Table \ref{Tab:lines}. For the fit to the CO(7--6) and \ci(2--1) lines we fix the line width to the mean value derived from the unblended lines. Their uncertainties include the variations of the line intensities for a fit where the line width is a free parameter.

\subsection{Observations of CO(3--2) and \cii\ }
We used the 7\,mm receivers of the Australia Telescope Compact Array (ATCA) to observe the CO(3--2) line (project ID: CX352).  Observations were carried out with the hybrid H214 array, which yields a beam size of 5-6$''$ at the observing frequency of 43.77\,GHz.  The line is detected with a SNR of 5.0 at a frequency and line width consistent with the ALMA derived redshift and line profiles.

In addition, we used the Atacama Pathfinder Experiment (APEX) to observe \cii\ at 240.57\,GHz.  The observations were carried out in 2016 April--May in good weather conditions with a precipitable water vapor content $<$1.5\,mm (project IDs E-296.A-5041B-2016 and M-097.F-0019-2016).  The observations were performed and the data processed as described by \citet{gullberg15}.  The \cii\ line is detected with a SNR of 4.3. From ALMA high spatial resolution observations of the \cii\ line (Marrone \etal\ in prep.; project ID: 2016.1.01293.S), we extract a \cii\ spectrum and flux, which are in good agreement with the APEX data. We adopt the ALMA \cii\ flux hereafter.

The line parameters derived from Gaussian fits to the data are given for both transitions in Table \ref{Tab:lines}; the spectra are shown in Figure \ref{Fig:spectra}.

\subsection{FIR dust continuum} \label{sec:photometry}
{Table \ref{Tab:lines} (right)} summarizes the dust continuum observations of SPT0311$-$58.
With seven broadband continuum detections between 3\,mm to 250\,\micron, the far-infrared spectral energy distribution (SED) of SPT0311$-$58 is thoroughly covered.

The SPT 1.4 and 2.0\,mm flux densities were extracted and deboosted as described by \citet{mocanu13}. 
We obtained a 870\,\micron\ map  with APEX/LABOCA (project ID: M-091.F-0031-2013).  The data were obtained, reduced, 
and the flux extracted following \citet{greve12}.  
Using {\it Herschel}/SPIRE, we obtained maps at 250\,\micron , 350\,\micron\ and 500\,\micron\ (project ID: DDT\_mstrande\_1).  The data were obtained and reduced as described by \citet{strandet16}.  

From our photometry, we derive an apparent far-infrared (FIR) luminosity (integrated between 40--120\,\micron\ rest) of L$_{{\rm FIR}}=4.1\pm\,0.7\,\times\,10^{13}$\,\lsol\ (see Figure \ref{Fig:SED-LVG}).

\floattable
\begin{deluxetable}{l c c c c | c c}
\tablecaption{Observed properties of SPT0311-58 \label{Tab:lines}}
\tablecolumns{5} 
\tablewidth{0pt} 
\tablehead{
\multicolumn{5}{c}{\bf Line properties} & \multicolumn{2}{c}{\bf Continuum properties}}
\startdata
\bf Transition	& \bf $\int\textit{SdV}$ 	& \bf $\textit{dV}^{a}$	& \bf $\textit{L}'$			& \bf $\textit{L}$  	& \bf Wavelength & \bf $\textit{S}_{\nu}$\\
 & \bf [Jy km/s]	& \bf [km/s] & \bf $\times$10$^{11}$\,[K\,km/s/pc$^2$]	 & \bf $\times$10$^{8}$\,[$L_{\odot}$] 	& \bf [\micron]	& \bf [mJy] \\ \hline
CO(3--2)	& 0.96 $\pm$ 0.15	& 790 $\pm$ 150	& 1.52 $\pm$ 0.24	& 2.01 $\pm$ 0.32 	& 3000	& 1.30 $\pm$ 0.05 	\\
CO(6--5)	& 2.10 $\pm$ 0.33	& 720 $\pm$ 140	& 0.83 $\pm$ 0.13	& 8.8 $\pm$ 1.4	& 2000	& 7.5 $\pm$ 1.3	\\
CO(7--6)	& 2.78 $\pm$ 0.80	& 750$^b$		& 0.81 $\pm$ 0.11	& 13.6 $\pm$ 1.8	& 1400	& 19.0 $\pm$ 4.2	\\
\ci(2--1)	& 1.29 $\pm$ 0.80	& 750$^b$		& 0.37 $\pm$ 0.10	& 6.4 $\pm$ 1.8	& 870 	& 32.0 $\pm$ 5.0	\\
\cii$_{\rm APEX}$ & 22.1 $\pm$ 5.1 & 890 $\pm$ 260 & 1.16 $\pm$ 0.27	& 254 $\pm$ 59 	& 500	& 52.0 $\pm$ 8.0	\\
\cii$_{\rm ALMA}$ & 25.88 $\pm$ 0.65 & 			& 1.36 $\pm$ 0.03 & 298.1 $\pm$ 7.5 	& 350  	& 38.0 $\pm$ 6.0	\\
 &  		 &			&  & & 250 	& 29.0 $\pm$ 8.0	\\
\enddata
\tablenotemark{$^{a}$ FWHM} \tablenotemark{$^b$ fixed from CO(3--2) and CO(6--5). }\\
\end{deluxetable}

\section{Characterizing the ISM in SPT0311$-$58}
\label{sec:LVG+SED-modeling}
\subsection{Source properties from high resolution imaging}
ALMA high spatial resolution imaging (angular resolution of 0.3 $\times$ 0.5$\arcsec$) of the \cii\  line in SPT0311$-$58 shows that the system consists of two galaxies in close proximity (Marrone \etal\ in prep.). Only the western source is significantly gravitationally magnified and this source dominates the apparent continuum luminosity ($>90$\% of the restframe 160\,\micron\ continuum flux density is emitted by the western source). 
In the following, we assume that the contribution from the eastern source is negligible and model the system as a single object, using the system magnification of $\mu = 1.9$ (Marrone \etal\ in prep.).

\begin{figure*}
\centering
\includegraphics[height=5.5cm]{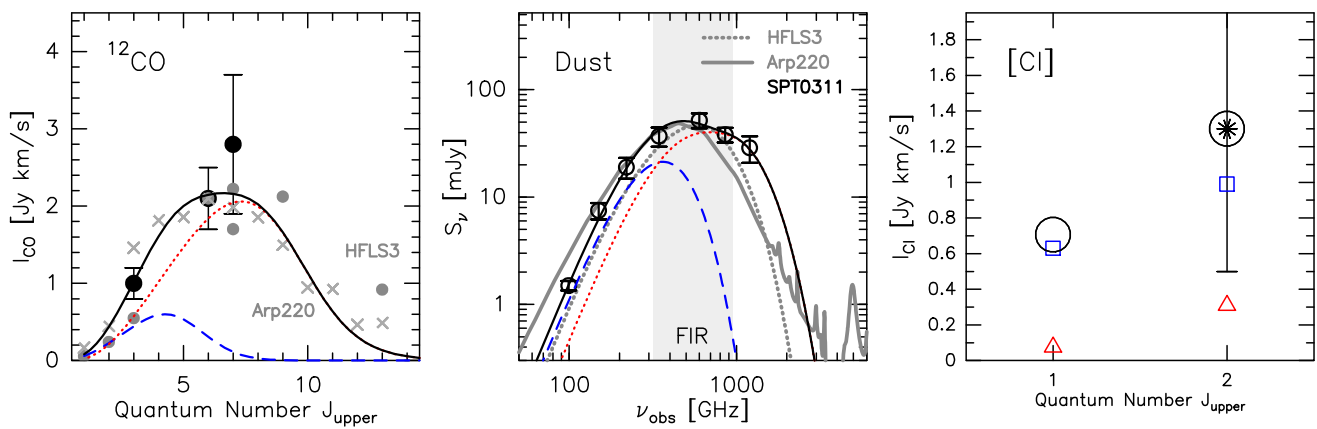}
\caption{\small Results of simultaneous fitting of the CO SLED (\emph{left}), SED (\emph{middle}) and \ci\ (\emph{right}) line flux. The CO line intensities are plotted as filled circles, the continuum fluxes as open circles and the \ci flux as an asterisk. The blue dashed line and squares show the cold component, the red dotted line and triangles show the warm component, the black solid line and circles (in right panel) show the combination of the two components. In grey is shown similar data for HFLS3 (dots and dotted line) and Arp220 (crosses and solid line), normalized to the CO(6--5) flux of SPT0311-58 for the CO-SLED and to the peak of the continuum SED of SPT0311-58 for the SED. The line fluxes and continuum properties fitted are listed in Table \ref{Tab:LVG-results}.
\emph{Left}: Two component CO-SLED.
\emph{Middle}:  Two component SED fitting based.
\emph{Right}: The contribution of each of the two components to the \ci(2--1) line and predictions for the \ci(1--0) line. 
}
	\label{Fig:SED-LVG}
\end{figure*}

\subsection{Radiative Transfer Models}
We use the FIR photometry and the line luminosities from Table \ref{Tab:lines} to simultaneously model the dust continuum, CO spectral line energy distribution (SLED), and the \ci(2--1) line following the radiative transfer calculation presented in \citet{weiss07}.  In this model, the background radiation field is set to the cosmic microwave background (CMB) for the dust and to the CMB plus the dust radiation field for the lines. The line and dust continuum emission are further linked via the gas column density in each component that introduces the turbulence line width as a free parameter in the calculation \citep[see Eq. 7 in][]{weiss07}. The gas column density calculated from the line emission together with the gas to dust mass ratio (GDMR) then determines the optical depth of the dust.

The calculations treat the dust and the kinetic temperature as independent parameters, but with the prior that the kinetic gas temperature has to be equal to or higher than the dust temperature.
Physically, this allows for additional sources of mechanical energy (e.g. shocks) in the ISM in addition to photo-electric heating. 

The chemical parameters in our model are the CO and \ci\ abundances relative to H$_2$ and the GDMR. 
We use a fixed CO abundance of $8 \times 10^{-5}$ relative to H$_2$ \citep{frerking82},
but keep the \ci\ abundance and the GDMR as free parameters.  For the frequency dependence of the dust absorption coefficient we adopt
$\kappa_{\rm d}(\nu)=0.04\,(\nu/250\,{\rm GHz})^{\beta}$ [m$^2$\,kg$^{-1}$]  \citep{krugel94}, which is in good agreement with $\kappa_{870\mu m} = 0.077$\,m$^2$\,kg$^{-1}$ used in other work \citep[see][and references therein]{spilker15}, for our best-fitting $\beta$.

Model solutions are calculated employing a Monte-Carlo Bees \citep{pham09} algorithm which randomly samples the parameter space and gives finer sampling for good solutions (as evaluated from a $\chi^{2}$ analysis for each model). In total, we sample $\sim 10^7$ models.  Parameter values and uncertainties were calculated using the probability-weighted mean of all solutions and the standard deviations.

\subsection{Model results}

\floattable
\begin{deluxetable}{c c c c c}
\tablecaption{ISM parameters of SPT0311-58 from the radiative transfer calculation \label{Tab:LVG-results}}
\tabletypesize{\footnotesize}
\tablecolumns{7} 
\tablewidth{0pt} 
\tablehead{
\colhead{\bf Parameter} & {\bf unit} & \colhead{\bf overall} 	& \colhead{\bf cold component}  			& \colhead{\bf warm component}
}
\startdata
Equivalent radius\,$^{a}$	& pc			& (4000 $\pm$ 1700)\,$\mu^{-1/2}$	& (3700 $\pm$ 1300)\,$\mu^{-1/2}$	&  (1500 $\pm$ 1200)\,$\mu^{-1/2}$ \\
$T_{\rm dust}$ 			& K              	&                       					& 36 $\pm$ 7                     			& 115 $\pm$ 54   				\\
$T_{\rm kin}$ 			& K               	&                       					& 58 $\pm$ 23                    			& 180 $\pm$ 51   				\\
log($n$(\hh)) 			& cm$^{-3}$      	&                       					& 3.7 $\pm$ 0.4                  			& 5.1 $\pm$ 1.9   				\\
$dv_{\rm turb}$ 		& \kms  		&                       					& 130 $\pm$ 17                  			& 100 $\pm$ 4 					\\
$\kappa_{\rm vir}\,^{b}$    &     			&                       					& 1.9 $\pm$ 1.9				& 3.1 $\pm$ 2.5				\\
GDMR                		&             		&  							& \multicolumn{2}{c}{110 $\pm$ 15$^d$}							\\
$\beta$ 				&			&							&\multicolumn{2}{c}{1.91 $\pm$ 0.05$^d$}						\\
\ci/[\hh]		   		&             		&  							& (6.0 $\pm$ 1.4)$\times 10^{-5}$ 	& (1.7 $\pm$ 2.3)$\times 10^{-5}$	\\
$M_{\text{dust}}$ 		& $M_{\odot}$	& (5.7 $\pm$ 0.8)$\times 10^9$\,$\mu^{-1}$ & ($5.2\pm0.7$)$\times 10^9$\,$\mu^{-1}$ & ($4.8\pm0.7$)$\times 10^8$\,$\mu^{-1}$ \\
$M_{\text{gas}}$ 		& $M_{\odot}$	& (6.3 $\pm$ 3.7)$\times 10^{11}$\,$\mu^{-1}$	& ($5.7\pm3.8$)$\times 10^{11}$\,$\mu^{-1}$	& ($5.3\pm3.8$)$\times 10^{10}$\,$\mu^{-1}$		\\
$\alpha_{\text{CO}}$		&  $M_{\odot}$/K\,\kms\,pc$^2$ & 4.8 $\pm$ 2.9	& 5.5 $\pm$ 4.0					& 3.1 $\pm$ 2.5					\\
$L_{\text{FIR}}$ 		& $L_{\odot}$	& (4.1 $\pm$ 0.7) $\times 10^{13}$\,$\mu^{-1}$	& $(1.2\pm1.1)\times 10^{13}$\,$\mu^{-1}$	& $(2.9\pm0.7) \times 10^{13}$\,$\mu^{-1}$	 	\\		
SFR$\,^{c}$ 			& $M_{\odot}\,$yr$^{-1}$	& $(4100\pm700$)\,$\mu^{-1}$	 &						&        						\\
$t_{\text{dep}}$ 		& Myr		& 150 $\pm$ 90				& 			        				&       						\\
\enddata
\tablecomments{The values here are apparent values. Intrinsic values can be calculated using $\mu = 1.9$.}

\tablenotemark{$^{a}$ $r_0={\rm D}_{\rm A}\,\sqrt{\Omega_{s}/\pi}$}

\tablenotemark{$^{b}$ $dv/dr$=$\kappa_{\rm vir}\times3.1\sqrt{n(\hh)/1e4}$; we calculate the velocity gradient for virialized clouds \citep[$\kappa_{vir}=1$,][]{goldsmith01} but also consider}\\ 
\tablenotemark{nonvirial, unbound motions  \citep[$\kappa_{vir}>1$,][]{greve09}.}

\tablenotemark{$^{c}$ using SFR=$10^{-10} \times L_{\rm FIR}$ based on a Chabrier initial mass function \citep{kennicutt98b,chabrier03}}

\tablenotemark{$^d$ Fitted in the radiative transfer calculation but set to be the same for both components.} 
\end{deluxetable}

Figure \ref{Fig:SED-LVG} shows the CO SLED, the continuum SED, and \ci\ flux density. From the figure, it is apparent that the dust continuum SED cannot be modeled with a single temperature modified blackbody, so we instead fit two components. Since we have no information on the high-$J$ CO transition, we use the shape of the CO SLED of Arp220 \citep{rosenberg15} and HFLS3 \citep{riechers13} as priors. With this choice, we compare the moderately excited CO SLED of Arp220 (see \citet{rosenberg15} for a comparison of Arp220 to other local ULIRGs) to the more extreme case of HFLS3 where the CO SLED stays high up to the $J_{\rm up}$=9 level (see Fig. \ref{Fig:SED-LVG}). The use of the priors mainly affects the parameters of the warm gas and therefore only has a small effect on our derived gas mass (see below).
Table \ref{Tab:LVG-results} lists the parameters obtained from the radiative transfer calculations for the Arp220 prior, not corrected for magnification. 

For both priors, the warm dust component dominates the peak of the CO SLED and the short wavelength part of the dust spectrum and therefore the FIR luminosity.
Its size is small compared to the cold gas with an area ratio of $\sim6$ 
($r_0=1.7\pm1.4$\,kpc where $r_0$ is the equivalent radius defined as r$_0={\rm D}_{\rm A}\,\sqrt{\Omega_{s}/\pi}$ \citep{weiss07})
for HFLS3 and slightly smaller for Arp220) which implies that the region of intense FIR continuum emission is significantly smaller than the overall gas distribution. Due to a lack of observations of CO transitions beyond (7--6), its properties are mainly driven by the assumed shape of the CO SLED for the high-$J$ transitions. But the models for both priors indicated consistently that the warm gas has a substantial density (of order 10$^5$\,cm$^{-3}$), a dust temperature of $\sim 100$\,K and a kinetic temperature in excess (but consistent within the errors) of the dust temperature (\tkin=180\,$\pm$\,50\,K when using Arp220 priors).

The cold dust component is required to fit the CO(3--2) and \ci\ line emission, and the long wavelength part of the dust SED. Due to its large extent and relatively high density (r$_0$=3.7\,$\pm$\,1.3\,kpc, log(n(H$_2$)=3.7\,$\pm$\,0.4)), it carries 
$\approx$\,90\% of the gas mass. The abundance of neutral carbon in this gas phase is \ci$/[\hh]=6.0\pm1.4\times10^{-5}$ in agreement with other estimates at high redshift and in nearby galaxies \citep[e.g.][and references therein]{weiss05}. For both priors, the cold gas dominates the CO(1--0) line luminosity. As for the warm gas, we find that the kinetic temperature is above the dust temperature (T$_{\text{dust}}$=36\,$\pm$\,7\,K, \tkin=58\,$\pm$\,23\,K), which may suggest that the ISM in SPT0311-58 experiences additional mechanical energy input e.g., via feedback from stellar winds or AGN driven outflows. This is also supported by the large turbulent line width of order 100 \kms\ and super-virial velocity gradients ($\kappa_{\rm vir} > 1$, see note b in Table \ref{Tab:LVG-results}) we find for both components and priors.

We use the kinematic parameters (dv$_{\rm turb}$ and $\kappa_{\rm vir}$)
together with the source size and the H$_2$ density for each component \citep[see Eq. 8 in][]{weiss07} to derive a total apparent gas mass of M$_{\rm gas}$=(6.3\,$\pm$\,3.7)\,$\times$\,$10^{11}$\msol\ (including a 36\% correction to account for the cosmic He abundance). For the HFLS3 prior, the gas mass is $\sim30$\% higher.

\section{Discussion}
\subsection{Gas mass conversion factor}
With the independent gas mass estimate from the radiative transfer models in-hand we can also derive the gas-to-dust mass ratio (GDMR) and the CO-to-H$_2$ conversion factor ($\alpha_{\text{CO}}$) for SPT0311$-$58. Since the CO(1--0) transition has not been observed, we use the flux density from the radiative transfer model which predicts I$_{\rm CO(1-0)}$=0.10\,$\pm$\,0.03 Jy km/s.
In our models, we assume that each gas component has the same GDMR and we find GDMR=110\,$\pm$\,15. Due to the different physical conditions in each gas component, there is a specific $\alpha_{\text{CO}}$ value for each component.  For the cold dust  component, we find $\alpha_{\text{CO}} = 5.5 \pm 4.0 $\,$M_{\odot}$(K\,km\,s$^{-1}$\,pc$^2)^{-1}$ 
and for the warm dust component $\alpha_{\text{CO}} = 3.1 \pm 2.5 $\,$M_{\odot}$(K\,km\,s$^{-1}$\,pc$^2)^{-1}$.  
Combining both gas components we find for SPT0311$-$58 $\alpha_{\text{CO}} = 4.8 \pm 2.9 $\,$M_{\odot}$(K\,km\,s$^{-1}$\,pc$^2)^{-1}$.  

When calculating gas masses for ULIRGs, a factor of $\alpha_{\text{CO}} = 0.8$\,$M_{\odot}$(K\,km\,s$^{-1}$\,pc$^2)^{-1}$ 
is typically assumed \citep{downes98}, significantly below our estimate. The difference can easily be explained by the much higher densities we find in both components compared to the models from \citet{downes98} in which most of the CO(1--0) luminosity arise from a diffuse inter-cloud medium. Since the bulk of the gas mass of this source is in the dense component, it is vital to include the higher-$J$ CO transitions in the calculation of $\alpha_{\text{CO}}$.

A similar two-component analysis was done for the broad absorption line quasar APM08279+5255 at $z=3.9$ \citep{weiss07}, where the dense component was found to dominate the CO(1--0) line by 70\%. They find a high conversion factor of $\alpha_{\text{CO}} \sim 6$\,$M_{\odot}$(K\,km\,s$^{-1}$\,pc$^2)^{-1}$, similar to what we find in the dense gas component. A similar reasoning for higher CO conversion factors owing to the presence of dense gas was put forward by \citet{papadopoulos12} based on the CO SLED in local (U)LIRGs.

\subsection{\cii}
\begin{figure}
\centering
\includegraphics[height=6.9cm,angle=0]{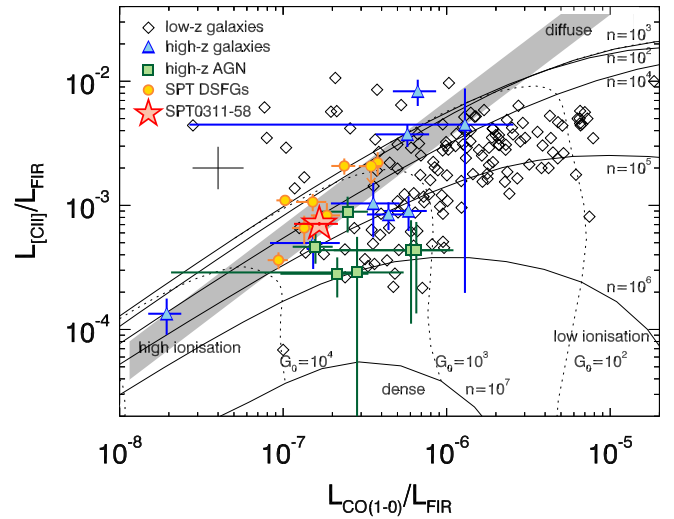}
	\caption{\small $L_{\rm \cii}$/$L_{\text{FIR}}$ vs $L_{\rm \cii}$/$L_{\text{CO(1--0)}}$ with PDR models and samples adapted from \citet{gullberg15}. The red star shows how SPT0311-58 falls within the larger SPT DSFG sample. The typical error bar for the low redshift sample is presented by the black cross.  
}
	\label{Fig:OIII-CII}
\end{figure}

From our \cii\ detection, we derive a $L_{\rm \cii}$/$L_{\text{FIR}}$ ratio of (7.3\,$\pm$\,0.1)$\times 10^{-4}$.  Figure \ref{Fig:OIII-CII} shows that this puts SPT0311$-$58 into the lower region of the $L_{\rm \cii}$/$L_{\text{FIR}}$ ratio observed in a larger sample of SPT-DSFGs \citep{gullberg15}. Similarly, low $L_{\rm \cii}$/$L_{\text{FIR}}$ ratios are found for the $z=6.3$ star-forming galaxy HFLS3 \citep{riechers13} and for the $z=7.1$ QSO host galaxy J1120+0641\citep{venemans12}. 

%

The $L_{\rm \cii}$/$L_{\text{CO(1--0)}}$ ratio in SPT0311$-$58 is similar to what is observed in the SPT sample \citep[4300\,$\pm$\,1300 compared to 5200\,$\pm$\,1800][]{gullberg15}, and HFLS3 \citep[$\sim$3000,][]{riechers13}. 
 This is consistent with the picture in which the \cii\ emission stems from the surface of dense clouds exposed to the strong UV field from the intense starburst in SPT0311$-$58 \citep{stacey10,gullberg15, spilker16}. 

The larger \cii\ deficit together with the decreasing $L_{\rm \cii}$/$L_{\text{CO(1--0)}}$ ratio of SPT0311$-$58 and other high redshift sources compared to local galaxies may be understood as a consequence of an increasing gas surface density \citep{narayanan17}:  the higher molecular gas surface density pushes the H$_{\rm I}$ + H$_2$ mass budget towards higher H$_2$ fractions. Since \cii\ mainly arises from the PDR zone associated with H$_{\rm I}$ and the outer H$_2$ layer, this effect reduces the size of the \cii\ emitting region and therefore the \cii\ line intensity. At the same time, the ratio of $L_{\rm \cii}$/$L_{\text{CO(1--0)}}$ will decrease due to an increase in the fraction of carbon locked in CO compared to \cii.

\subsection{Concluding remarks}
Both our radiative transfer model and fine structure line results indicate that SPT0311-58 resembles typical DSFGs, just at $z \sim 7$. 
This is also supported by its extreme SFR surface density of $\Sigma_{\rm SFR} \sim 600$\,M$_{\odot}$yr$^{-1}$kpc$^{-2}$ (derived using the size of the warm gas component that dominates the FIR luminosity) which approaches the modeled values for radiation pressure limited starbursts \citep[$10^3$\,M$_{\odot}$yr$^{-1}$kpc$^{-2}$][]{thompson05} and is comparable to what is found in other starburst like Arp220, HFLS3 and other SPT-DSFGs \citep{scoville03, riechers13,spilker16}.
Future observations of this source will explore its spatial structure, physical conditions, formation history, and chemical evolution in great detail as it is one of very few massive galaxies known at $z \sim 7$

\acknowledgments 											 %
MLS was supported for this research through a stipend from the International Max Planck Research School (IMPRS) for Astronomy and Astrophysics at the Universities of Bonn and Cologne. 
MA acknowledges partial support from FONDECYT through grant 1140099. 
JDV, DPM, KCL, JSS and SJ acknowledge support from the U.S. National Science Foundation under grant No. AST-1312950. 
BG acknowledges support from the ERC Advanced Investigator programme DUSTYGAL 321334.
The Flatiron Institute is supported by the Simons Foundation.
Partial support for DN was provided by NSF AST-1009452, AST-1445357, NASA HST AR-13906.001 from the Space Telescope Science Institute, which is operated by the Association of University for Research in Astronomy, Incorporated, under NASA Contract  NAS5-26555, and a Cottrell College Science Award, awarded by the Research Corporation for Science Advancement. 
ALMA is a partnership of ESO (representing its member states), NSF (USA) and NINS (Japan), together with NRC (Canada) and NSC and ASIAA (Taiwan), in cooperation with the Republic of Chile. The Joint ALMA Observatory is operated by ESO, AUI/NRAO and NAOJ.
The National Radio Astronomy Observatory is a facility of the National Science Foundation operated under cooperative agreement by Associated Universities, Inc.
{\it Herschel} is a European Space Agency Cornerstone Mission with significant participation by NASA. 
APEX is a collaboration between the Max-Planck-Institut f\"{u}r Radioastronomie, the European Southern Observatory, and the Onsala Space Observatory.
The Australia Telescope is funded by the Commonwealth of Australia for operation as a National Facility managed by CSIRO.
The SPT is supported by the National Science Foundation through grant PLR-1248097, with partial support through PHY-1125897, the Kavli Foundation and the Gordon and Betty Moore Foundation grant GBMF 947.



\end{document}